\documentclass[a4paper,11pt]{article}
\usepackage{pos}
\usepackage{tikz-feynman}
\usepackage{mathtools}
\usepackage{hyperref}

\title{Improving the electromagnetic form factor of the pion at large $Q^2$ using the Feynman-Hellmann theorem}
\ShortTitle{Improving the Electromagnetic Form Factor of the Pion}

\author[a]{K.~U.~Can}
\author[a]{J.~A.~Crawford}
\author[b]{R.~Horsley}
\author*[a]{J.~J.~McKee}
\author[c]{P.~E.~L.~Rakow}
\author*[a]{I.~van~Schalkwyk}
\author[d]{G.~Schierholz}
\author[e]{H.~St\"{u}ben}
\author[a]{R.~D.~Young}
\author{J.~M.~Zanotti${^a}$ (QCDSF Collaboration)}

\affiliation[a]{CSSM, Department of Physics, School of Physics, Chemistry and Earth Sciences, The University of Adelaide, Adelaide SA 5005, Australia.} 
\affiliation[b]{School of Physics and Astronomy, University of Edinburgh, Edinburgh EH9 3JZ, UK.} 
\affiliation[c]{Theoretical Physics Division, Department of Mathematical Sciences, University of Liverpool, Liverpool L69 3BX, UK.} 
\affiliation[d]{Deutsches Elektronen-Synchrotron DESY, Notkestr. 85, 22607 Hamburg, Germany.} \affiliation[e]{Regionales Rechenzentrum, Universit\"{a}t Hamburg, 20146 Hamburg, Germany.}

\emailAdd{jordan.mckee@adelaide.edu.au}
\emailAdd{ian.vanschalkwyk@adelaide.edu.au}

\abstract{
At large momentum transfer, it becomes increasingly difficult to access the form factor of the pion $F_\pi(Q^2)$ using lattice QCD simulations. Two of the limiting factors include the increased computational cost of adding more statistics to overcome gauge noise, as well as suppressed overlap with the ground state of the boosted pion. Here we apply two noise reduction techniques, all-mode averaging (AMA) and momentum smearing, to the computation of $F_\pi(Q^2)$ at high momentum transfers using the Feynman-Hellmann (FH) theorem. First, we show that all-mode averaging by itself produces good improvement compared to previous results, at an equal computational cost. We also implement a momentum smearing technique to further reduce statistical uncertainties. In contrast to conventional smearing approaches, our Feynman-Hellmann method requires combining back-to-back momentum states, and hence we adapt a version of smearing involving a superposition of back-to-back smearing operations. This method is then implemented to compute $F_\pi(Q^2)$ at $Q^2 = 6.6 \;\mathrm{GeV^2}$, demonstrating good improvement over the regular smeared counterpart. Finally both all-mode averaging and momentum smearing are combined to determine $F_\pi(Q^2)$ at $Q^2 = 6.6 \;\mathrm{GeV^2}$ showing an excellent preliminary improvement over previous calculations. 
}

\FullConference{The XVIth Quark Confinement and the Hadron Spectrum Conference (QCHSC24)\\
 19-24 August, 2024\\
 Cairns Convention Centre, Cairns, Queensland, Australia\\}

\begin{document}
\maketitle

\section{Introduction}
The complicated internal structure of hadrons provide valuable insight into the nature of quantum chromodynamics (QCD). Of particular interest are the electromagnetic form factors of hadrons which describe the elastic scattering of electrons off the hadronic system.

The pion is of particular interest as the lightest hadron in the QCD spectrum. The form factor of the pion $F_\pi(Q^2)$ provides valuable insight into the transition of QCD from the non-perturbative to perturbative regimes at high $Q^2$.

Probing these structures experimentally is quite challenging at higher momentum transfers, $Q^2$, due to the natural decay of the pion. Jefferson Lab (JLab) have determined $F_\pi(Q^2)$ up to $Q^2 = 2.45\;\mathrm{GeV^2}$, utilising the secondary mechanism of pion electro-production off of nucleons \cite{JeffersonLab:2008jve}. The JLab12 experimental program is currently underway to extend these measurements up to $6\;\mathrm{GeV^2}$ \cite{Dudek:2012vr}. In the future, it is anticipated that the future electron-ion Collider \cite{Arrington:2021biu} and Electron-ion collider in China \cite{Anderle:2021wcy} will be capable of probing the range $Q^2 \sim 9-40\;\mathrm{GeV^2}$.

With experimental results currently limited, lattice QCD calculations provide a benchmark for $F_\pi(Q^2)$. As lattice calculations have probed towards higher $Q^2$ regions, the statistical uncertainty of the results have been found to increase as a consequence of gauge field noise and contamination from excited states \cite{Lepage:1989hd}. In anticipation of future experimental results probing into higher $Q^2$ regions, these statistical restrictions have to be overcome.

In previous investigations using lattice QCD, the accuracy of the form factor calculations have been limited by gauge field noise, especially as the momentum of the state increases, as observed in our previous analysis in Ref.~\cite{QCDSF:2017ssq}. To overcome this limitation, various noise reduction techniques can be utilised. This work will consider the methods of all-mode averaging (AMA) \cite{Shintani:2014vja} and momentum smearing \cite{Bali:2016lva}. AMA serves to provide an increase in statistics when computing correlation functions without significantly increasing the computational cost. Momentum smearing improves the signal-to-noise ratio (SNR) of correlators by improving ground state overlap with quarks at high momentum and suppressing overlap with excited states.  

A recent calculation of the form factor of the pion has been able to determine it up to $10\;\mathrm{GeV^2}$ by Ref. \cite{Ding:2024lfj}, extending to larger momentum transfers than previous lattice QCD calculations, with the application of the AMA technique to overcome statistical limitations.

This work aims to extend upon the evaluations of Ref. \cite{QCDSF:2017ssq}, which utilise the FH technique to determine the form factor. By implementing AMA alongside momentum smearing to improve the precision and accuracy of the $F_\pi(Q^2)$ calculation without the usual associated increase in computational cost. Additionally, this combination of techniques alongside the FH technique can provide improvements in determining other quantities of interest, such as the Compton amplitude of hadrons \cite{Can:2020sxc}.

\section{Form Factor}
\begin{figure}[h!]
\centering
\begin{tikzpicture}[scale=1.5, transform shape, every path/.style={line width=0.3mm}]
            \begin{feynman}
                \vertex (a);
                \vertex [above right=of a] (b) [blob] {}; 
                \vertex [below left=1.35cm and 1.35cm of b] (a1) {\(\pi\)};
                \vertex [above left =2cm of b] (f1) {\(\gamma^*\)};
                \vertex [right=2cm of b] (c) {\(\pi\)};
            
                \diagram* {
                  (a1) -- [fermion, momentum={\(p\)}] (b) [blob] -- [fermion, momentum={\(p'\)}] (c),
                  (f1) -- [boson, momentum={\(q\)}] (b),
                };
            \end{feynman}
\end{tikzpicture}
\caption{Insertion of an electromagnetic current into a pion}
\label{fig:Scatter}
\end{figure}
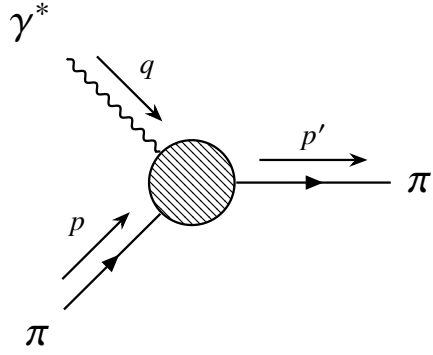

In order to probe the internal structure of hadrons, scattering experiments are performed. In the case of elastically scattering a charged lepton, such as an electron, off of a pion, a virtual photon is exchanged as depicted in Fig.~\ref{fig:Scatter}. 
This interaction will provide a transfer in 4-momentum of $Q^2 = -q^2 = - (p'-p)^2$ which changes the pion's 4-momentum from $p$ to $p'$ without changing the state of the pion. The matrix element corresponding to the insertion of a photon current as an external field $\mathcal{V}_\mu(z) = \overline{q}(z)\gamma_\mu q(z)$ into the pion is 
\begin{align}
    \langle \pi(\textbf{p}')|\mathcal{V}_\mu(0)|\pi(\textbf{p})\rangle = -i(p'+p)_\mu F_\pi(Q^2)
\end{align}
which describes the electromagnetic structure of the pion with respect to a single Lorentz invariant function $F_\pi(Q^2)$. This form factor encapsulates the distribution of the electric charge within a pion.

For the purposes of this investigation, the incident-normal Breit frame $\textbf{p}' = \textbf{p}$ is utilised. This provides only a transfer of momentum without the transfer of energy, while also simplifying the FH relations. Note that since this work is performed on a Euclidean lattice, the momentum transfer and form factor decompositions are in Euclidean space-time.

\subsection{Analytical representations of $F_\pi(Q^2)$}
QCD lends itself well to perturbative solutions at high $Q^2 \gg \Lambda_\text{QCD}$, where $\Lambda_\text{QCD} = 338\; \mathrm{MeV}$  is the typical QCD scale, given for $N_f = 3$ quark flavours \cite{FlavourLatticeAveragingGroupFLAG:2024oxs}. The perturbative QCD (pQCD) evaluation, according to Ref.~\cite{Chang:2013nia}, of the form factor is
\begin{align} \label{eq:pQCD}
    F_\pi(Q^2) \xrightarrow{\text{large}\ Q^2} \frac{16\pi}{Q^2}\alpha_s(Q^2)f_\pi^2
\end{align}
where $f_\pi$ is the pion decay constant and $\alpha_s(Q^2)$ is the renormalized strong coupling.
At low energies where hadronic degrees of freedom become important, $F_\pi(Q^2)$ is modelled using the Vector Meson Dominance (VMD) model \cite{OConnell:1995fwv}. The VMD model assumes the exchanged photon is dominated by a vector meson at lower energies. This model provides $F_\pi(Q^2)$ of the form
\begin{align}
    F_\pi(Q^2) \approx \frac{1}{1 + Q^2/m_\rho^2},
\end{align}
which serves as a general guideline for the form factor at low energies, but is expected to breakdown as the system transitions to the perturbative regime. The scale at which this transition occurs and the form of this transition are not well known.

\section{Lattice Methodology}

We consider $2+1$ quark mass degenerate flavours of stout-link non-perturbatively $O(a)$-improved clover fermions \cite{Cundy:2009yy} on a $32^3\times64$ lattice with lattice spacing $a\sim 0.074$~fm. The quark masses are tuned to the SU(3)-symmetric point where $u$, $d$ and $s$ quarks are approximately set to the physical flavour-singlet mass $\overline{m} = (2m_l+m_s)/3$ \cite{Bietenholz:2010jr} 
corresponding to $m_\pi\approx470$~MeV \cite{Bietenholz:2011qq}.

\subsection{Two-point Correlation Functions} \label{sec:2pt}

The momentum projected hadron two-point correlation functions that are calculated on the lattice are defined by

\begin{align} \label{eq:2pt}
    G_{\chi {\chi^\dagger}}(\mathbf{p};t) = \sum_x e^{-i \mathbf{p} \cdot \mathbf{x}} \langle \Omega | \chi (t, \mathbf{x}) \chi^\dagger (0, \mathbf{0})|\Omega \rangle.
\end{align}
Here, the following pseudoscalar interpolating operator is used for states of the pion
\begin{align}
    \chi^{\text{pion}}(x) = u^\dagger(x) \gamma_5 d(x).
\end{align}
The right-hand side of Eq.~(\ref{eq:2pt}) may be expressed as a sum of all possible intermediate states with definite momenta
\begin{align}
   G_{\chi {\chi^\dagger}}(\mathbf{p};t) = \sum_n \frac{e^{-E_n (\mathbf{p}) t}}{2E_n (\mathbf{p})} \langle\Omega| \chi (0) | n, \mathbf{p} \rangle \langle n, \mathbf{p} | \chi^\dagger (0) | \Omega \rangle,
\end{align}
where $n$ labels the energy level with $n=1$ being the ground state. The interpolating operator creates and annihilates all states with the same quantum numbers as a pion. This means the correlation function is contaminated by excited states at small Euclidean times. The degree to which the $n^\text{th}$ excited state is exponentially suppressed depends on its energy, $E_n(\mathbf{p})$, and is given by, $e^{-E_n (\mathbf{p}) t}$, as well as the ratio of the overlap $n^\text{th}$ excited state with respect to the overlap of the ground state, $| \langle n, \mathbf{p} | \bar{\chi} (0) | \Omega \rangle|^2 /| \langle 1, \mathbf{p} | \bar{\chi} (0) | \Omega \rangle|^2$. To improve excited state suppression, quark smearing is employed, which improves the overlap of the interpolating operators with the ground state while suppressing their overlap with the excited states, especially the low-lying excited states. We build on this idea in Sec.~\ref{sec:momSmear} with the concept of momentum smearing for boosted states. For this work, gauge invariant Jacobi smearing is employed except where otherwise specified.

To extract the energy from the two-point correlator, a single exponential fit is used at sufficiently large Euclidean times.

\subsection{Feynman-Hellmann Method} \label{sec:FH}
In quantum mechanics, the FH theorem relates the change in energy of a system with respect to some perturbation to the expectation value of the derivative of the Hamiltonian with respect to the same perturbation. This method may be extended to lattice QCD and it has shown to produce comparable results to using more conventional three-point function methods as shown in Refs.~\cite{QCDSF:2012mkm,CSSM:2014uyt,QCDSF:2017ssq}. The FH theorem in quantum mechanics gives the relation
\begin{align}
    \frac{\partial E(\lambda)}{\partial \lambda}  = \langle \psi_n | \frac{\partial H(\lambda)}{\partial \lambda} | \psi_n \rangle,
\end{align}
where $H(\lambda)$ is the perturbed Hamiltonian and $E(\lambda)$ the energy eigenvalue of the state $\psi_n$ with respect to the real parameter $\lambda$. An equivalent theorem exists for lattice QCD as shown in Refs.~\cite{QCDSF:2012mkm,CSSM:2014uyt}. In Ref.~\cite{QCDSF:2017ssq} it was shown that the electric form factor of the nucleon can be isolated through a modification of the action using the real parameter $\lambda$,

\begin{align} \label{eq:action}
    S_{\text{QCD}} \rightarrow S_\text{QCD}(\lambda) &= S_{\text{QCD}} + \lambda\int d^4z(e^{iq\cdot z}+e^{-iq\cdot z})\mathcal{V}_4(z).
\end{align}
The matrix element of interest, $\langle \pi (\mathbf{p}') | \mathcal{V}_4(0) | \pi (\mathbf{p}) \rangle$, is only non-zero in the off forward condition ($\mathbf{q} \neq 0$) if $\mathbf{p'}=\mathbf{p}\pm\mathbf{q}$. The FH implementation shown in Eq.~(\ref{eq:action}) introduces a two-fold degeneracy in momentum space. In this case the off-forward matrix elements will only survive for sets of degenerate states satisfying $\mathbf{p'}=\mathbf{p}\pm\mathbf{q}$, which is realised in Breit frame kinematics where $\mathbf{p'}=\mathbf{-p}$. The eigenstates of the perturbed system are given by a linear combination of the momentum states, which we construct by using the average of correlators that are projected to definite momenta $\mathbf{p}$ and $\mathbf{-p}$.
Hence, for insertions of the temporal components of the vector current respectively we get

\begin{align}
    \frac{\partial E}{\partial \lambda} \bigg|_{\lambda = 0} =-i\frac{[p'+p]_4}{2E_\pi(\mathbf{p})}F_\pi(Q^2) = F_\pi(Q^2).
\end{align}
The energy of the perturbed correlator, $E_\lambda$, may then be calculated to extract the energy shift and hence apply the FH theorem. The perturbed two-point correlator for the nucleon at large Euclidean time takes the form,
\begin{align}\label{eq:corrE}
    G^N_\lambda(\mathbf{p},\mathbf{q}, t) \xrightarrow{t \gg 1} A_\lambda (\mathbf{p},\mathbf{q})e^{-E_\lambda^N(\mathbf{p}, \mathbf{q})t},
\end{align}
where $E_\lambda^N$ is the perturbed energy of the ground state nucleon and $A_\lambda$ is the corresponding overlap factor. For a nucleon, we may then construct a ratio $R_\lambda$ of perturbed and unperturbed correlators to obtain an energy shift $\Delta E$,

\begin{align} \label{eq:ratio}
    R_\lambda = \frac{G^N_\lambda(\mathbf{p}, \mathbf{q} ,t)}{G^N_0 (\mathbf{p}, t)} \xrightarrow{t \gg 1} \frac{A_\lambda(\mathbf{p}, \mathbf{q})}{A_0(\mathbf{p})}e^{-\Delta E t}.
\end{align}
For the pion however, the pion correlator takes the following form
\begin{align}
    G_\lambda(\mathbf{p}, \mathbf{q}, t)\xrightarrow{t, T-t \gg1} A_\lambda(\mathbf{p}, \mathbf{q}) \cosh \bigg[E_\lambda^N(\mathbf{p}, \mathbf{q}) \bigg( t - \frac{T}{2} \bigg)\bigg],
\end{align}
where $T$ is the temporal extent of the lattice. constructing a ratio in the same way as Eq.~(\ref{eq:ratio}) for the pion becomes more difficult due to the hyperbolic cosine functions which do not simplify easily to give $\Delta E$. Instead of forming a ratio, we extract the perturbed and unperturbed energies from the ground state using two-point correlators of the form,
\begin{align}
    G^\pi_\lambda (\mathbf{p}, t) = (A_0 + \Delta A_\lambda )e^{-(E^\pi_0 + \Delta E^\pi)t} + (A_0 - \Delta A_\lambda )e^{-(E_0^\pi - \Delta E^\pi)(T-t)} 
\end{align}
where $T$ is the temporal extent of the lattice. The change in sign for the energy shift of the backwards state compared to the forward state is dictated by the time reversal properties of the electromagnetic current. 
The shifted energy may be Taylor expanded as
\begin{align}\label{eq:taylor}
    E_{\lambda}(\mathbf{p, q}) = E_0^\pi(\mathbf{p}) \pm \Delta E^\pi(\mathbf{p,q}) = E_0(\mathbf{p})\pm \lambda\frac{\partial E(\mathbf{p, q})}{\partial \lambda} + \frac{\lambda^2}{2!}\frac{\partial^2 E(\mathbf{p, q})}{\partial^2 \lambda} + \mathcal{O}(\lambda^3).
\end{align}
The unperturbed energy $E_0^\pi(\mathbf{p})$ may then be subtracted from Eq.~\ref{eq:taylor}. A quadratic fit can then be used to fit to the terms which are linear and quadratic in $\lambda$, at which point we may isolate the first order energy shift of interest,

\begin{align}
    \Delta E_\lambda(\mathbf{p,q}) = \lambda\frac{\partial E(\mathbf{p, q})}{\partial \lambda}.
\end{align}
By selecting sufficiently small values of $\lambda$, any $\mathcal{O}(\lambda^3)$ and higher terms are suppressed.

\subsection{All-Mode Averaging}
AMA is a class of covariant approximation averaging (CAA) techniques developed by Shintani \textit{et. al} \cite{Shintani:2014vja}. AMA takes advantage of a relaxed stopping condition when implementing the conjugate gradient method for computing fermion propagators. This takes into account the contributions of all eigenmodes of the Dirac matrix, as opposed to its predecessor, low-mode averaging, which only retains low-lying modes.

\subsubsection{Covariant Approximation Averaging}
In our case, CAA will be applied to the computation of two-point correlation functions, but more broadly can be applied to any observable of interest. Two-point correlation functions are constructed using quark propagators which are determined via inversion of the Dirac matrix. CAA utilises strict correlators $G_\text{strict}$ which are calculated utilising a double-precision accuracy in the Dirac matrix inversion solver tolerance, while approximate solves for $G_\text{appx}$ vary in construction based on the method being utilised but generally have a lower precision. The method utilised to produce $G_\text{appx}$ defines the techniques of low-mode averaging and AMA. CAA reduces the computational cost of calculating correlators $G(t)$ by utilising $N_\text{appx}$ approximations $G_\text{appx}$ which are corrected using $N_\text{strict}$ strict correlators $G_\text{strict}$ computed in the same conditions i.e. source/sink locations, gauge configurations and quark source smearing. The correction factor is determined by averaging over the difference between a set of strict and approximate solves with matching source locations
\begin{align}\label{eq:Corr}
    G_\text{corr} = \frac{1}{N_\text{strict}}\sum_{i=1}^{N_\text{strict}}(G_\text{strict}^i - G_\text{appx}^i).
\end{align}
$G_\text{corr}$ is then utilised to approximate solves an improved correlator for each $G_\text{appx}^j$ where $j\in[N_\text{strict}+1,N_\text{appx}]$ with
\begin{align}
    G_\text{imp}^j = G_\text{appx}^j + G_\text{corr}.
\end{align} 
The improved correlators are then averaged over with the strict correlators to produce
\begin{align}
    G_\text{AMA} = \frac{1}{N_\text{appx}}\bigg(\sum_{i=1}^{N_\text{strict}}G_\text{strict}^i + \sum_{j=N_\text{strict}+1}^{N_\text{appx}}G_\text{imp}^j \bigg).
\end{align} 
Provided $G_\text{appx}$ is cheaper to compute than $G_\text{strict}$, CAA techniques provide an improvement in statistics when computing observables by taking advantage of the reduced computational cost of computing approximate correlators. Instead of calculating additional strict correlators to increase the number of statistics, the equivalent amount of computational time is put towards computing approximate correlators. Since $G_\text{appx}$ is cheaper to produce than $G_\text{strict}$, more sources can be produced for an equivalent computational cost. 

\subsubsection{AMA Implementation}
In practice, AMA is implemented by computing approximations to the correlation function $G_{\text{appx}}$ where the inverter algorithm is provided with a relaxed stopping condition. This will provide an approximate correlator with a lower precision ($\varepsilon = 10^{-3}$) then the usual strict solutions ($\varepsilon = 10^{-12}$). This lower precision provides the $G_{\text{appx}}$ with a reduced computation time compared to their strict counterparts as desired. This provides a cheaper way to increase the statistics of an analysis. $G_{\text{appx}}$ is then corrected using a correction factor $G_\text{corr}$ produced in accordance with Eq.~(\ref{eq:Corr}). The improved correlator $G_\text{AMA}$ is then utilised to carry on with the analysis to extract the desired observables.

\begin{figure}[h]
    \centering
    \includegraphics[width=1\linewidth]{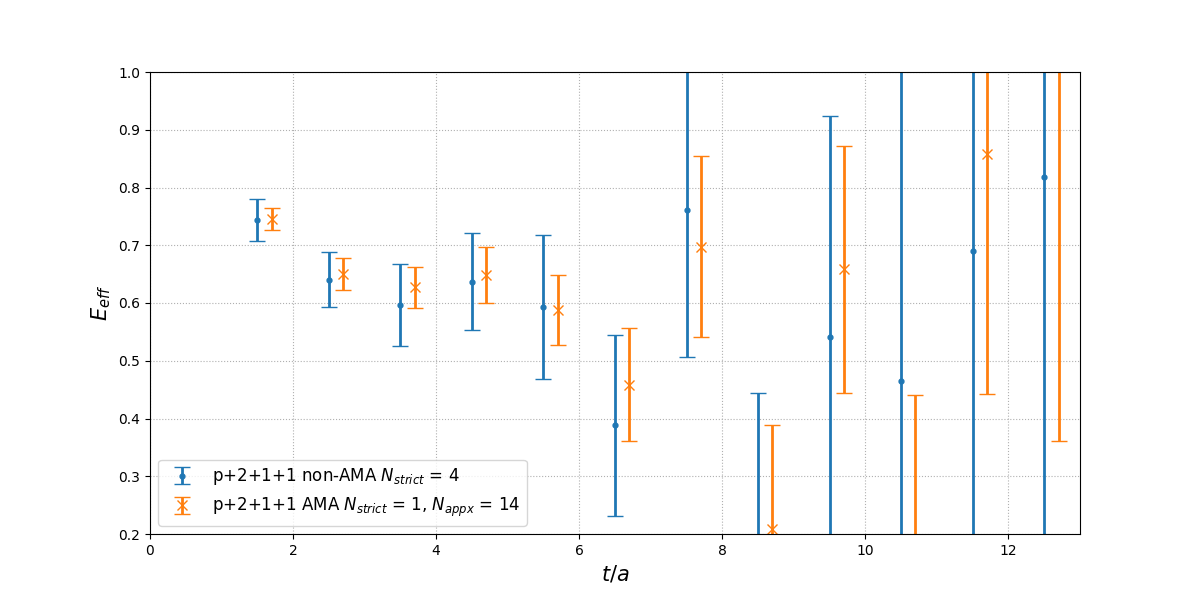}
    \caption{Equal cost comparison of the effective mass of a non-AMA correlator utilising $4$ strict sources compared to an AMA correlator utilising $N_\text{strict} = 1$ and $N_\text{appx} = 14$}
    \label{fig:AMA_meff_Comp}
\end{figure}

To demonstrate the statistical improvement of AMA, Fig. \ref{fig:AMA_meff_Comp} provides a comparison between the effective masses of a non-AMA and AMA correlator for a pion boosted to momentum $\mathbf{p} = (2,1,1)\frac{2\pi}{L}$,  utilising a roughly equal computational cost. The effective energy is determined using the equation
\begin{align}
    E_\text{eff} = \ln{\frac{G(t)}{G(t+1)}}
\end{align}
which provides the ground state energy when applied to Eq.~(\ref{eq:corrE}).

The benefits of utilising AMA are noticeable in Fig. \ref{fig:AMA_meff_Comp} by the increased precision of the AMA effective mass compared to the non-AMA case. This increased precision indicates a more precise determination of the correlator energy states can be produced when fitting directly to the AMA correlator over the non-AMA case.

\subsubsection{Combining AMA with Feynman-Hellmann}\label{sec:FHAMA}
As discussed in Sec.~\ref{sec:FH}, the FH theorem gives a relation between the energy shift and the perturbed system. In applying a perturbation of some order $\mathcal{O}(10^{-p})$ we require the precision $\varepsilon$ of the approximated correlator to be smaller ($\varepsilon < 10^{-p}$) to be able to resolve the corresponding energy shift. 

\begin{figure}[h]
    \centering
    \includegraphics[width=1\linewidth]{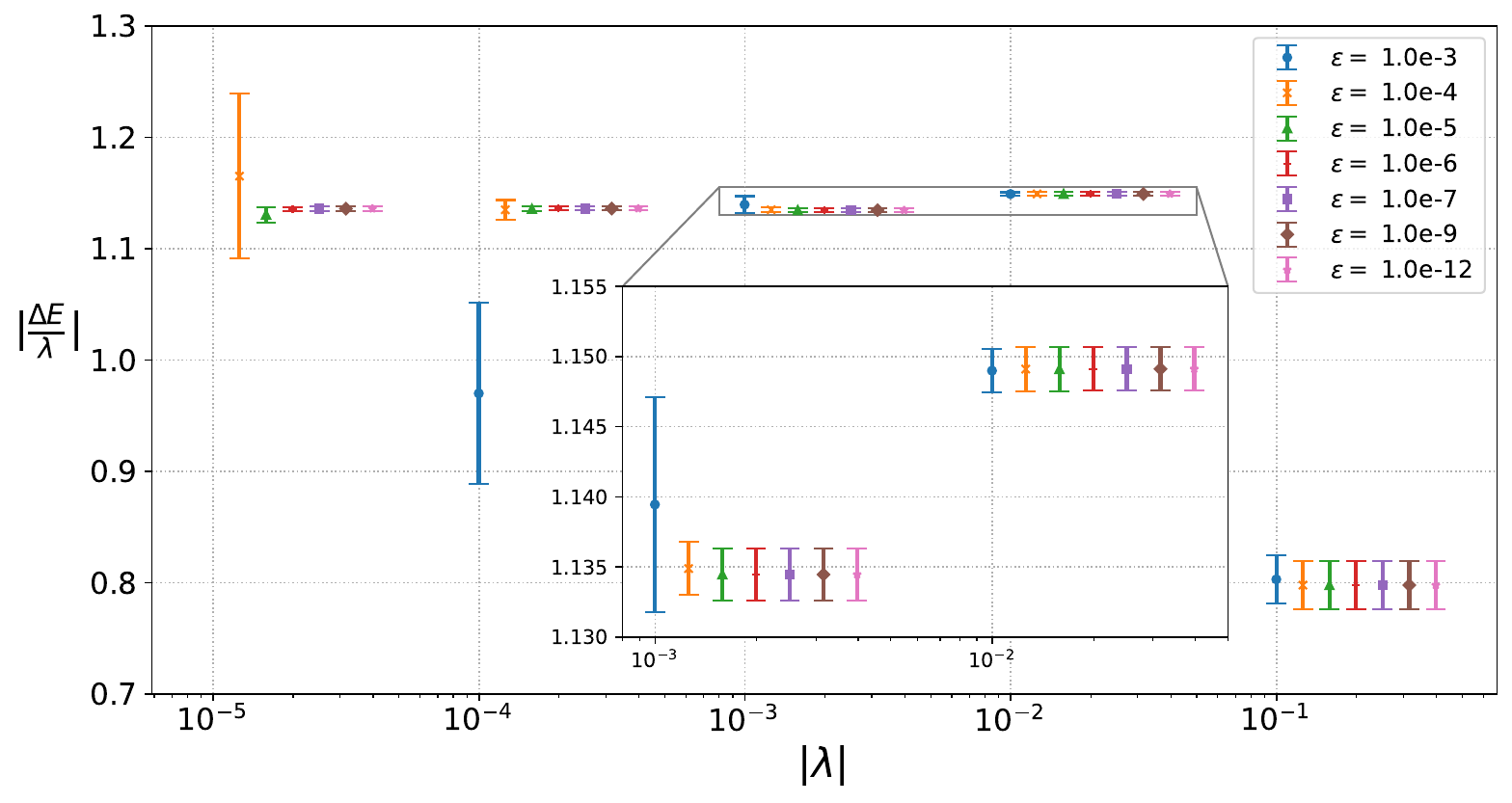}
    \caption{A comparison of the determined energy shifts using FH for $F_\pi(Q^2)$ using various precisions and perturbation strengths. Matching $\lambda$ values are shifted for visibility and the $\lambda = 10^{-5}$ blue point is above the graph. An enlarged view is provided for the larger $\lambda$ cases for clarity in identifying the signal behavior.}
    \label{fig:PrecLambAnalysis}
\end{figure}

The effect of AMA with Feynman-Hellmann is determined by analysing the resulting energy shifts corresponding to various perturbation strengths $\lambda \in \{10^{-5},10^{-4},10^{-3},10^{-2},10^{-1}\}$ for various precisions $\varepsilon$. The results of this analysis are presented in Fig. \ref{fig:PrecLambAnalysis}, which demonstrates how the signal breaks down when $\lambda \lesssim \varepsilon$ with growing uncertainties and a deviation from accepted values at higher precisions. 

This is of particular importance to identify appropriate values of $\lambda$ to utilise in our analysis. To overcome this loss of information, larger $\lambda$ values need to be utilised relative to the desired precision. Additionally, since the form factor corresponds to the linear energy shift, low values of $\lambda$ are required to suppress the higher-order $\lambda$ contributions. To compromise between both influences, the choices of $\lambda$ are $\lambda = 0, 0.005, \pm 0.05$ for a desired $G_\text{appx}$ precision of $\varepsilon = 10^{-3}$ in accordance with Fig. \ref{fig:PrecLambAnalysis}. The presence of any quadratic contaminants will be included within the fit to the energy shifts and removed from the analysis.

Through further analysis, it is also evident that this observed effect is $Q^2$ dependent in accordance with the magnitude of the resultant energy shift. A higher $|\mathbf{q}|$ corresponds to larger energy shifts which provides a more reliable signal in the smaller $\lambda$, smaller precision cases.

For future applications of AMA alongside the Feynman-Hellmann theorem it is important to investigate this effect on a case by case basis to avoid a loss of signal. With these limitations in mind we continue forward with careful attention made to avoid a loss of signal with the choices of $\lambda$ and inverter precision.

\subsection{Momentum Smearing} \label{sec:momSmear}
As mentioned in Sec.~\ref{sec:2pt}, another method to improve ground state overlap is to utilise freedom in interpolator choice and use interpolators resembling the ground state wave function of particles. Unfortunately, because states with higher momenta have faster decaying two- and three-point correlators, excited state suppression becomes less effective for these hadrons leading to a degraded SNR. Bali \textit{et al.} \cite{Bali:2016lva} propose a method called momentum smearing, which is designed to improve the ground state overlap of hadrons with high momenta while adding little additional computational cost.
\subsubsection{Quark Smearing}
Smearing algorithms take a point-like source (or sink) and extend them to a Gaussian-shaped object, acknowledging that hadrons are not point-like objects. To get an idea of how this may work, consider the following general smearing algorithm,
\begin{align} \label{eq:smear}
    (Fq)_\mathbf{x} = \sum_{\mathbf{y} \in (a\mathbb{Z})^d} f_{\mathbf{x}-\mathbf{y}} G_{\mathbf{xy}} q_{\mathbf{y}},
\end{align}
where $d$ is the number of spatial dimensions, for this work $d=3$, $f$ is a scalar function, $G$ the gauge covariant transporter and $q_{\mathbf{y}}$ the quark field at spatial position $\mathbf{y}$ and $(Fq)_\mathbf{x}$ is the smeared quark field at the spatial position $\mathbf{x}$. We recognize that Eq.~(\ref{eq:2pt}) takes the form of a convolution. Hence, taking the Fourier transform of Eq.~(\ref{eq:2pt}) we get a product in momentum space
\begin{align}
    \sum_{\mathbf{x} \in (a\mathbb{Z})^d} e^{i \mathbf{p} \cdot \mathbf{x}} (Fq)_{\mathbf{x}} = \Tilde{f}(\mathbf{p)}\Tilde{q}_{\mathbf{p}},
\end{align}
where $\Tilde{q_\mathbf{p}}$ is the momentum space quark field. Letting $f$ be a Gaussian in position space with amplitude $f_0$ and width $\sigma$,
\begin{align}
    f_{\mathbf{x}-\mathbf{y}} = f_\mathbf{0}\, \mathrm{exp} \bigg(-\frac{|\mathbf{x}-\mathbf{y}|}{2 \sigma^2} \bigg),
\end{align}
the corresponding $\Tilde{f}$ in momentum space will then be a Gaussian centered around $\mathbf{p}=0$,
\begin{align}
    \Tilde{f}(\mathbf{p}) = \Tilde{f}(\mathbf{0}) \, \mathrm{exp} \bigg(-\frac{\sigma^2\mathbf{p}^2}{2} \bigg).
\end{align}
This smearing procedure will then only maximize the overlap for hadrons at zero momentum, however, making it less efficient for boosted states.
\subsubsection{Momentum Smearing Implementation}
In Ref~\cite{Bali:2016lva}, Bali \textit{et al.} proposed that by centering the momentum space Gaussian around some momentum of interest $\mathbf{k}$ and performing another Fourier transform, we need only modify the position space smearing function as follows
\begin{align}
    \Tilde{f}(\mathbf{p}) \rightarrow \Tilde{f}(\mathbf{p-k}) \implies f_\mathbf{z} \rightarrow e^{i\mathbf{k}\cdot \mathbf{z}}f_\mathbf{z}.
\end{align}
Applying this to our general smearing equation we get
\begin{align}
        (Fq)_x = \sum_{\mathbf{y} \in (a\mathbb{Z})^d} e^{-i\mathbf{k} \cdot (\mathbf{x} - \mathbf{y})} f_{\mathbf{x}-\mathbf{y}} G_{\mathbf{xy}} q_{\mathbf{y}}.
\end{align}
This modification, which may be added to existing smearing algorithms such as Wuppertal or Jacobi 
smearing, is what Bali \textit{et al.} term `\textit{momentum smearing}' in Ref.~\cite{Bali:2016lva}. To implement momentum smearing on existing smearing algorithms, it is sufficient to substitute the gauge link located at $\mathbf{x}$, pointing in the direction $\hat{j}$, by, $U_{\mathbf{x}, j} \rightarrow U_{\mathbf{x}, j} e^{i \mathbf{k} \cdot \hat{\mathbf{j}}}$. Since smearing is applied at the quark level, for hadron momenta $\mathbf{p}$ we choose $ \mathbf{k} = \mathbf{p}/2,\ \mathbf{p}/3$ for mesons and baryons respectively.

\subsubsection{Superposition Momentum Smearing}
Momentum smearing is effective at increasing the ground state overlap of a hadron with momentum $\mathbf{p}$. However, momentum smearing also introduces an anisotropy in the imaginary component of the smeared wave function which causes the overlap of a hadron with momentum $-\mathbf{p}$ to be significantly degraded, even compared with its $\mathbf{k=0}$ smeared counterpart. However, as discussed in Sec.~\ref{sec:FH}, for the pion form factor $F_\pi(Q^2)$ our implementation of the FH theorem requires that the eigenstate of the perturbed system be constructed from the linear combination of projected definite momenta states with momentum $\mathbf{p}$ and $\mathbf{-p}$. We propose a method which uses the superposition of two sources, one smeared with $\mathbf{k}$ and the other with $-\mathbf{k}$. The sink may then be smeared in either $\mathbf{k}$, $-\mathbf{k}$, or as a superposition of both.
For the Breit frame this provides an improvement for both the positive and negative momenta $\mathbf{p}$ and $-\mathbf{p}$.

\section{Results}

\subsection{All-Mode Averaging Only}\label{sec:AMA}
To provide a considerable computational advantage, a precision of $\varepsilon = 10^{-3}$ for the approximations was chosen which provides $1\; C_\text{strict}\equiv 4.6\; C_\text{appx}$ in our case. Based on our investigation in Sec \ref{sec:FHAMA}, to avoid the degradation of the signal while maintaining a high computational advantage, larger perturbations of $\lambda = 0.005, \pm 0.05$ are used. Anticipating the contamination of higher order energy shifts, the choice of $\pm 0.05$ is made to minimise the quadratic contributions to better isolate the linear shift as required.

For comparison, the benchmark for the amount of computational time utilised was chosen to follow a previous analysis performed in Ref. \cite{QCDSF:2017ssq}. This previous analysis utilised the same set of gauge configurations with a total of $N_\text{conf} = 1700$ configurations utilised with perturbations $\lambda = +10^{-4},-10^{-5}$, and a single strict solve ($\varepsilon = 10^{-12}$) only.

In implementing AMA, we would like to convert the computational time utilised across the many configurations into the computation of multiple sources. To quantify the computational cost, the formula
\begin{align}
    \text{Cost} = N_\text{conf}\bigg(N_\text{strict} + \frac{N_\text{appx}}{R}\bigg)(1+N_\lambda)
\end{align}
was utilised, where $N_\lambda$ is the number of different non-zero $\lambda$ perturbations applied. The factor $R$ is the ratio of the computation times of $C_\text{strict}$ to $C_\text{appx}$, in our case $R \approx 4.6$ i.e. one $ C_\text{strict} \equiv 4.6\; C_\text{sloppy}$

\begin{figure}[h]
    \centering
    \includegraphics[width=1\linewidth]{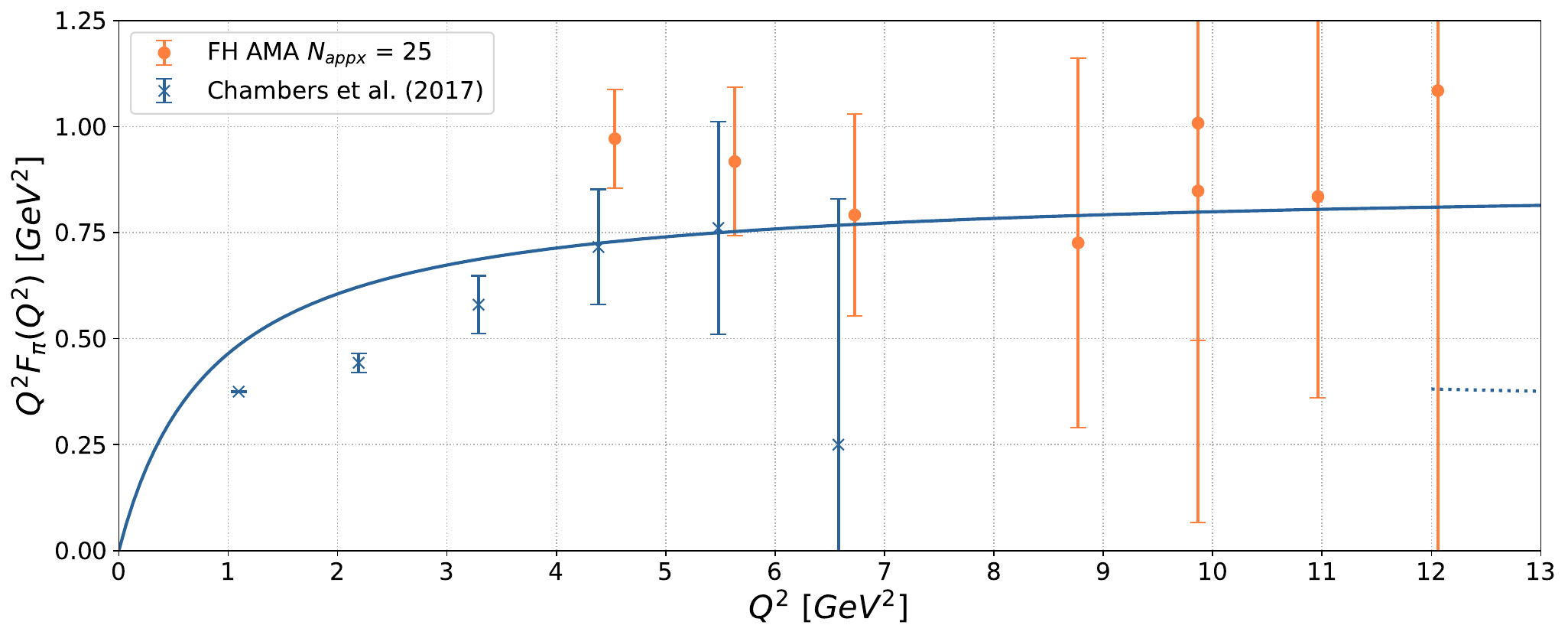}
    \caption{Comparison between $Q^2 F_\pi(Q^2)$ computed with AMA (orange points with $Q^2<7$ are shifted for clarity) for a roughly equal computation time comparison with results from Chambers \textit{et al.} Ref.~\cite{QCDSF:2017ssq} (blue). VMD prediction shown for $m_\rho = 932\;\mathrm{MeV}$ (solid blue line) and pQCD prediction, Eq.~(\ref{eq:pQCD}), shown for $f_\pi = 175\;\mathrm{MeV}$ \cite{QCDSF:2017ssq} and $N_f = 3$ (dotted blue line).}
    \label{fig:AMAFF25}
\end{figure}

This provides a convenient way to convert between different ways the computational cost is divided among the relevant features of our computations for the analysis performed on the same gauge configurations. We have settled on using $N_\text{conf} = 200$ for the comparison, which alongside the $N_\lambda = 3$ and $N_\text{strict} = 1$ provides a total of $N_\text{appx} = 25$ for an equal computation comparison with the simulation performed in Ref.~\cite{QCDSF:2017ssq}. Applying these choices across a range of $Q^2$ values up to $Q^2\approx 12\;\mathrm{GeV^2}$ provides the results in Fig. \ref{fig:AMAFF25}.

This shows an improvement as the reduction in the statistical uncertainty in $F_\pi(Q^2)$ for $Q^2$ values previously difficult to sufficiently isolate. Additionally an improvement is noticed at higher $8 < Q^2 < 11\;\mathrm{GeV^2}$ with $F_\pi(Q^2)$ entirely isolated to be non-negative. Pushing this further with additional computational time applied would allow the uncertainty of these points to be further reduced.

While AMA further improves the precision of the determination of $F_\pi(Q^2)$ by providing an increase in statistics, it is evident from Fig. \ref{fig:AMA_meff_Comp} that the correlators at earlier times still suffer from a contamination by excited states at higher $Q^2$. This contamination at higher $Q^2$ corresponds to the higher pion boosts required in accordance with the Breit frame kinematics required. The ground state energy of the system is more difficult to isolate as a result of this excited state contamination at higher momenta where the signal is lost to noise before ground state saturation is achieved. Momentum smearing provides a way to overcome these limitations by better representing the ground states of these boosted systems to reduce this excited state contamination.

\subsection{Momentum Smearing Only}

We begin with a direct comparison of three baryon effective mass plots that are constructed from quarks smeared with regular Jacobi smearing, momentum smearing with $\mathbf{k} = \mathbf{p}/3$, and a superposition of sources that have been momentum smeared with $\mathbf{k}$ and $-\mathbf{k}$. Two baryon momenta are considered, $\mathbf{p}_1 = \frac{2\pi}{L} (-3,0,0)$ and $\mathbf{p}_2 = \frac{2\pi}{L}(3,0,0)$. Hence, for momentum smearing we will choose $\mathbf{k}_1 = \frac{2\pi}{L}(-1, 0, 0)$ and $\mathbf{k}_2 = \frac{2\pi}{L}(1, 0, 0)$ to see the best improvements. For this comparison we use 1000 measurements.

\begin{figure}[h]
    \centering
    \includegraphics[width=1.0\linewidth]{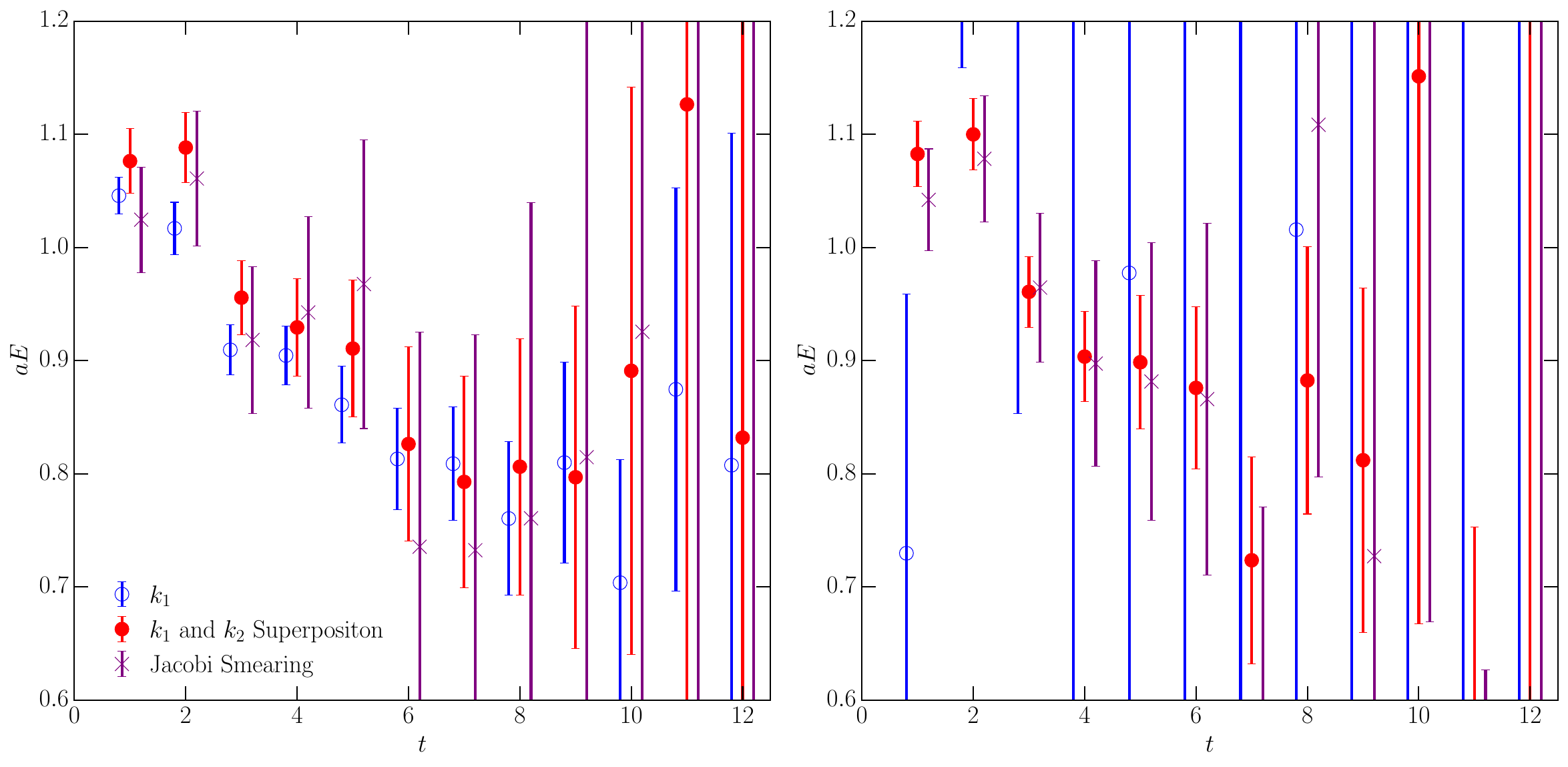}
    \caption{Comparison between effective mass plots for baryons with momentum $\mathbf{p}_1=\frac{2\pi}{L}(-3,0,0)$ (left) and $\mathbf{p}_2=\frac{2\pi}{L}(3,0,0)$ (right). Each plot shows effective energy of the baryon when smeared with Jacobi smearing (purple crosses), momentum smearing with $\mathbf{k}_1 = \frac{2\pi}{L}(-1,0,0)$ (blue circles) and superposition of both $\mathbf{k}_1$ and $\mathbf{k}_2$ (red dots).}
    \label{fig:2ptComp}
\end{figure}

Fig.~\ref{fig:2ptComp} shows the effective mass of two baryons with momentum $\mathbf{p}_1$ and $\mathbf{p}_2$ on the left and right respectively, with Jacobi smearing, momentum smearing with $\mathbf{k}_1$ and the superposition of momentum smeared sources with $\mathbf{k}_1$ and $\mathbf{k}_2$. The left plot shows that at $|\mathbf{p}_1| = |\mathbf{p}_2| = 3\big(\frac{2\pi}{L}\big)$ Jacobi smearing (purple crosses) yields a relatively high amount of noise. The aligned momentum smeared results (blue circles) on the left plot shows excellent improvement over the Jacobi smeared results (purple crosses). We also see on the left plot that the superposition of two momentum smeared sources results (red circles) also show good improvement over the Jacobi smeared results (purple crosses) but not as good as the aligned momentum smeared results (blue circles). On the right plot, the regular Jacobi smeared results (purple crosses) remain relatively unchanged as expected. However, the anti-aligned momentum smeared results (blue circles) on the right plot have a significantly degraded SNR. superposition of momentum smeared sources. The superposition of momentum smeared sources (red circles) on the right hand plot, however, shows significant improvement over both the anti-aligned momentum smeared results and the Jacobi smeared results.

\begin{figure}[h!]
    \centering
    \includegraphics[width=1.0\linewidth]{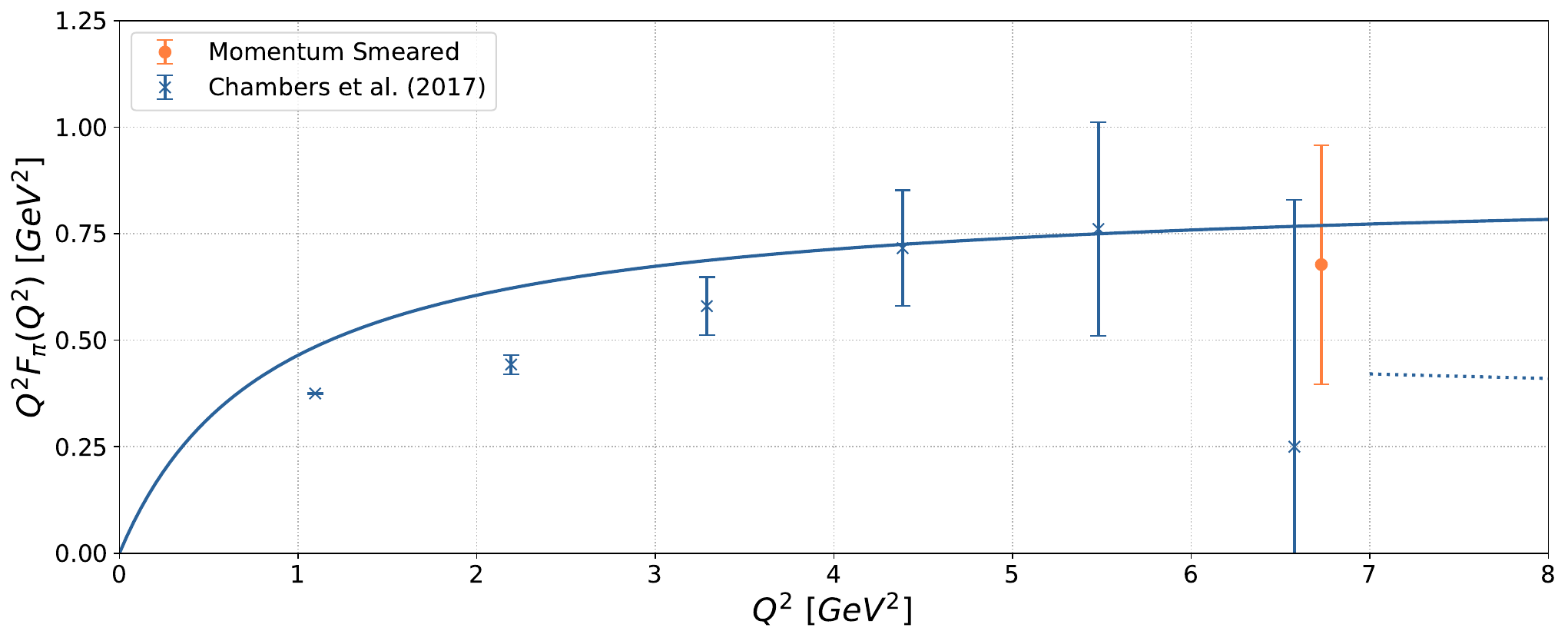}
    \caption{Comparison between $Q^2F_\pi(Q^2)$ calculated using regular Jacobi Smearing \cite{QCDSF:2017ssq} (blue) and using momentum smearing (orange), which has been shifted for clarity, with VMD shown for $m_\rho = 932\, \mathrm{MeV}$ (solid blue line) and pQCD prediction, Eq.~(\ref{eq:pQCD}), shown for $f_\pi = 175\, \mathrm{MeV}$ \cite{QCDSF:2017ssq} and $N_f = 3$ (dotted blue line)}
    \label{fig:MSFF}
\end{figure}

As improvement is seen in both the $\mathbf{p}_1$ and $\mathbf{p}_2$ states, we can use momentum smearing to calculate the pion form factor using the Feynman-Hellman technique in the Breit frame. A preliminary calculation of the pion form factor, $F_\pi(Q^2)$ is performed at $Q^2=6.6\,\mathrm{GeV^2}$ using $200$ configurations and is shown in Fig.~\ref{fig:MSFF}. The reduction in uncertainty and consistency with lower $Q^2$ points, as well as the VMD result, shows that there is considerable improvement over results obtained using standard smearing \cite{QCDSF:2017ssq}, with lower computational cost. More work running additional configurations is required to push calculations to higher $Q^2$.

\subsection{Combined results}
Provided that momentum smearing and AMA are applied to different steps in the computation of hadronic correlators, it is feasible that both methods can be utilised together. In practice, momentum smearing would improve the overlap with the ground state at higher momenta providing cleaner results, while AMA would increase the statistics to overcome the gauge noise. This should provide the ability to probe the behaviour of $F_\pi(Q^2)$ at higher $Q^2$ than previously achieved.

\begin{figure}[h!]
    \centering
    \includegraphics[width=1\linewidth]{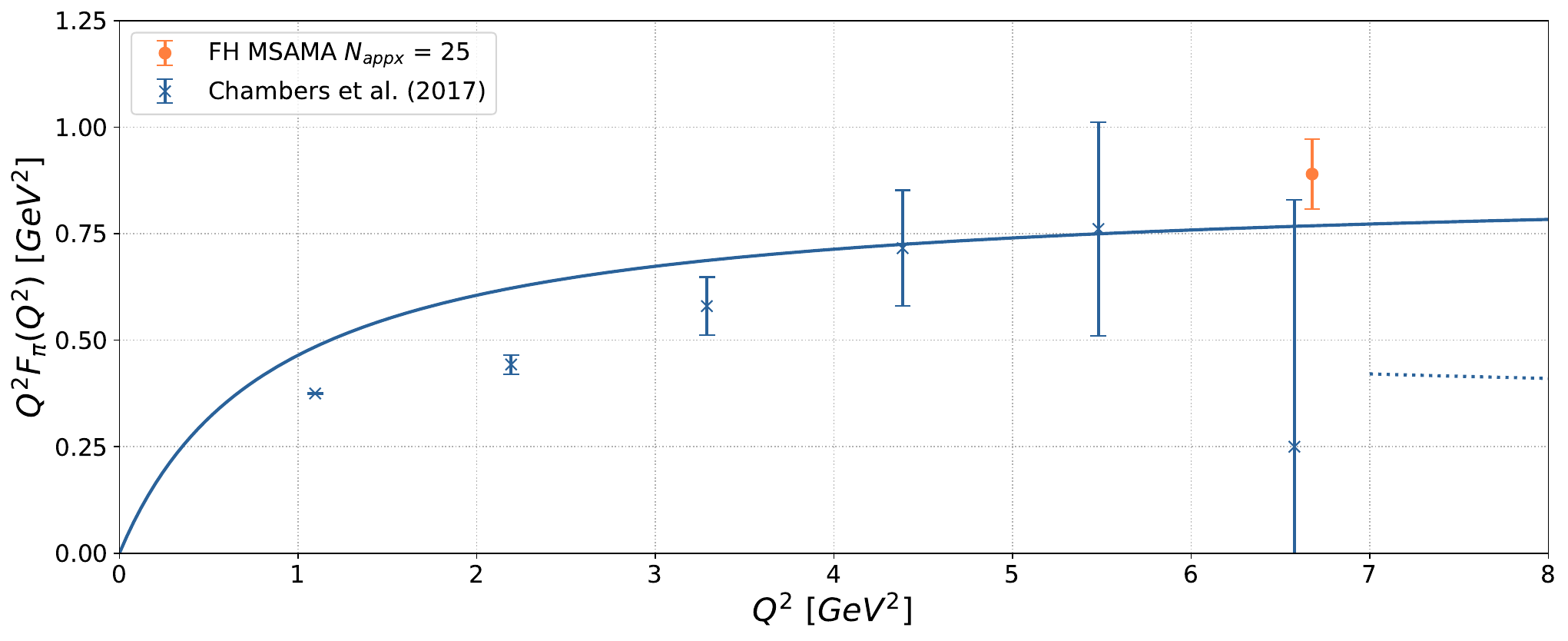}
    \caption{Pion form factor calculated using momentum smearing and AMA (MSAMA) (orange), which has ben shifted for clarity, plotted with Chambers ~\textit{et al.} 2017 Ref.~\cite{QCDSF:2017ssq} (blue crosses). VMD prediction shown at $m_\rho = 932 \mathrm{MeV}$ (blue line) as well as pQCD prediction, Eq.~(\ref{eq:pQCD}), for $f_\pi = 175\;\mathrm{MeV}$ \cite{QCDSF:2017ssq} with $N_f = 3$ (blue dotted line)}
    \label{fig:MSAMAFF25}
\end{figure}

We show our preliminary results from a preliminary calculation at $Q^2 = 6.6 \mathrm{GeV}^2$ shown in Fig.~\ref{fig:MSAMAFF25}, 
this shows a good improvement in the precision in the determined value of $F_\pi(Q^2)$ where the signal was previously lost. The combination of both methods yields a determination of $F_\pi(Q^2)$ which has a higher precision than either the use of momentum smearing or AMA individually. This demonstrates the combined methods are capable of more precisely determining $F_\pi(Q^2)$ at higher momenta than our previous analysis in Ref. \cite{QCDSF:2017ssq} with the same computational cost. Not only are we able to extract a non-zero value for the form factor where the signal was previously lost, we are now able to considerably improve its statistical precision and accuracy. The combined run is conducted with  $N_{\text{strict}} = 1$, $N_{\text{sloppy}} = 25$ and $200$ configurations.

\section{Conclusion}
We have presented results for two noise reduction techniques, AMA and momentum smearing. When utilising AMA, an improvement in the precision of the pion form factor is seen compared to our previous results \cite{QCDSF:2017ssq}. This calculation improves on previous results by clearly isolating the point at $Q^2 = 6.6\;\mathrm{GeV^2}$.

We have also shown that an improvement in the SNR of the two-point correlators of boosted baryons can be achieved when momentum smearing is applied in a desired direction. It was also shown that momentum smearing provides improvement to the SNR when a superposition of two sources which are smeared in opposite directions is taken, which is required for the pion form factor due to our Feynman-Hellmann implementation. These results led to the calculation of the pion form factor at $Q^2 = 6.6\;\mathrm{GeV^2}$ which also showed promising improvement, providing a more reliable value while roughly halving the uncertainty despite the relatively low number of gauge configurations.

Combining momentum smearing and all-mode averaging methods has provided major improvements for the determination of $F_\pi(Q^2)$ at $Q^2 = 6.6\;\mathrm{GeV^2}$, roughly decreasing the uncertainty by a factor of $6$. This provides an encouraging outlook for pushing the determination of $F_\pi(Q^2)$ using lattice QCD and the Feynman-Hellmann theorem to higher $Q^2$.

\newpage
\section*{Acknowledgments}
The numerical configuration generation (using the BQCD lattice QCD program \cite{Haar:2017ubh}) and data analysis (using the Chroma software library \cite{Edwards:2004sx}) was carried out on the Extreme Scaling Service (DiRAC, EPCC, Edinburgh, UK), the Data Intensive Service (DiRAC, CSD3, Cambridge, UK), the Gauss Centre for Supercomputing (NIC, J\"{u}lich, Germany), the NHR Alliance (Germany), and resources provided by the NCI National Facility in Canberra, Australia (supported by the Australian Commonwealth Government), the Pawsey Supercomputing Centre (supported by the Australian Commonwealth Government and the Government of Western Australia), and the Phoenix HPC service (University of Adelaide). RH is supported by STFC through grant ST/X000494/1. PELR is supported in part by the STFC under contract ST/G00062X/1. GS is supported by DFG Grant SCHI 179/8-1. JAC, JJM and IVS are supported by an Australian Government Research Training Program (RTP) Scholarship. KUC, RDY and JMZ are supported by the Australian Research Council grants DP190100297, DP220103098, and DP240102839. For the purpose of open access, the authors have applied a Creative Commons Attribution (CC BY) licence to any Author Accepted Manuscript version arising from this submission.


\begin{thebibliography}{99}
\bibitem{JeffersonLab:2008jve}
G.~M.~Huber \textit{et al.},
\textit{Charged pion form-factor between $Q^2 = 0.60\;\mathrm{GeV^2}$ and $2.45\;\mathrm{GeV^2}$. II. Determination of, and results for, the pion form-factor},
\href{https://journals.aps.org/prc/abstract/10.1103/PhysRevC.78.045203}{Phys. Rev. C \textbf{78} (2008) 045203,
doi:10.1103/PhysRevC.78.045203},
[\href{https://arxiv.org/abs/0809.3052}{arXiv:0809.3052 [nucl-ex]}].

\bibitem{Dudek:2012vr}
J.~Dudek, R.~Ent, R.~Essig, K.~S.~Kumar, C.~Meyer, R.~D.~McKeown, Z.~E.~Meziani, G.~A.~Miller, M.~Pennington, D.~Richards, L.~Weinstein and G.~Young,
\textit{Physics Opportunities with the 12 GeV Upgrade at Jefferson Lab}
\href{https://link.springer.com/article/10.1140/epja/i2012-12187-1}{Eur. Phys. J. A \textbf{48} (2012), 187
doi:10.1140/epja/i2012-12187-1}
[\href{https://arxiv.org/abs/1208.1244}{arXiv:1208.1244 [hep-ex]}].

\bibitem{Arrington:2021biu}
J.~Arrington, C.~A.~Gayoso, P.~C.~Barry, V.~Berdnikov, D.~Binosi, L.~Chang, M.~Diefenthaler, M.~Ding, R.~Ent and T.~Frederico, \textit{et al.}
\textit{Revealing the structure of light pseudoscalar mesons at the electron\textendash{}ion collider}
\href{https://www.researchgate.net/publication/350727588_Revealing_the_structure_of_light_pseudoscalar_mesons_at_the_electron-ion_collider}{J. Phys. G \textbf{48} (2021) 7, 075106
doi:10.1088/1361-6471/abf5c3}
[\href{https://arxiv.org/abs/2102.11788}{arXiv:2102.11788 [nucl-ex]}].

\bibitem{Anderle:2021wcy}
D.~P.~Anderle, V.~Bertone, X.~Cao, L.~Chang, N.~Chang, G.~Chen, X.~Chen, Z.~Chen, Z.~Cui and L.~Dai, \textit{et al.}
\textit{Electron-ion collider in China}
\href{https://link.springer.com/article/10.1007/s11467-021-1062-0}{Front. Phys. (Beijing) \textbf{16} (2021) 6, 64701
doi:10.1007/s11467-021-1062-0}
[\href{https://arxiv.org/abs/2102.09222}{arXiv:2102.09222 [nucl-ex]}].

\bibitem{Lepage:1989hd}
G.~P.~Lepage,
\textit{The Analysis of Algorithms for Lattice Field Theory},
\href{https://cds.cern.ch/record/206049}{CLNS-89-971}.

\bibitem{QCDSF:2017ssq}
A.~J.~Chambers, J.~Dragos, R.~Horsley, Y.~Nakamura, H.~Perlt, D.~Pleiter, P.~E.~L.~Rakow, G.~Schierholz, A.~Schiller, K.~Somfleth, H.~Stüben, R.~D.~Young and J.~Zanotti,
\textit{Electromagnetic form factors at large momenta from lattice QCD}
\href{https://journals.aps.org/prd/abstract/10.1103/PhysRevD.96.114509}{Phys. Rev. D \textbf{96} (2017) 11, 114509
doi:10.1103/PhysRevD.96.114509}
[\href{https://arxiv.org/abs/1702.01513}{arXiv:1702.01513 [hep-lat]}].

\bibitem{Shintani:2014vja}
E.~Shintani, R.~Arthur, T.~Blum, T.~Izubuchi, C.~Jung and C.~Lehner,
\textit{Covariant approximation averaging},
\href{https://journals.aps.org/prd/abstract/10.1103/PhysRevD.91.114511}{Phys. Rev. D \textbf{91} (2015) 11, 114511
doi:10.1103/PhysRevD.91.114511},
[\href{https://arxiv.org/abs/1402.0244}{arXiv:1402.0244 [hep-lat]}].

\bibitem{Bali:2016lva}
G.~S.~Bali, B.~Lang, B.~U.~Musch and A.~Sch\"afer,
\textit{Novel quark smearing for hadrons with high momenta in lattice QCD},
\href{https://journals.aps.org/prd/abstract/10.1103/PhysRevD.93.094515}{Phys. Rev. D \textbf{93} (2016) 9, 094515
doi:10.1103/PhysRevD.93.094515},
[\href{https://arxiv.org/abs/1602.05525}{arXiv:1602.05525 [hep-lat]}].

\bibitem{Ding:2024lfj}
H.~T.~Ding, X.~Gao, A.~D.~Hanlon, S.~Mukherjee, P.~Petreczky, Q.~Shi, S.~Syritsyn, R.~Zhang and Y.~Zhao,
\textit{QCD Predictions for Meson Electromagnetic Form Factors at High Momenta: Testing Factorization in Exclusive Processes, stout-link smearing},
\href{https://journals.aps.org/prl/abstract/10.1103/PhysRevLett.133.181902}{Phys. Rev. Lett. \textbf{133} (2024) 18, 181902
doi:10.1103/PhysRevLett.133.181902},
[\href{https://arxiv.org/abs/2404.04412}{arXiv:2404.04412 [hep-lat]}].

\bibitem{Can:2020sxc}
K.~U.~Can, A.~Hannaford-Gunn, R.~Horsley, Y.~Nakamura, H.~Perlt, P.~E.~L.~Rakow, G.~Schierholz, K.~Y.~Somfleth, H.~St\"uben and R.~D.~Young, J.~M.~Zanotti,
\textit{Lattice QCD evaluation of the Compton amplitude employing the Feynman-Hellmann theorem},
\href{https://journals.aps.org/prd/abstract/10.1103/PhysRevD.102.114505}{Phys. Rev. D \textbf{102} (2020), 114505
doi:10.1103/PhysRevD.102.114505},
[\href{https://arxiv.org/abs/2007.01523}{arXiv:2007.01523 [hep-lat]}].


\bibitem{FlavourLatticeAveragingGroupFLAG:2024oxs}
Y.~Aoki \textit{et al.}
\textit{FLAG Review 2024},
[\href{https://arxiv.org/abs/2411.04268}{arXiv:2411.04268 [hep-lat]}].

\bibitem{Chang:2013nia}
L.~Chang, I.~C.~Clo\"et, C.~D.~Roberts, S.~M.~Schmidt and P.~C.~Tandy,
\textit{Pion electromagnetic form factor at spacelike momenta},
\href{https://journals.aps.org/prl/abstract/10.1103/PhysRevLett.111.141802}{Phys. Rev. Lett. \textbf{111} (2013) 14, 141802
doi:10.1103/PhysRevLett.111.141802},
[\href{https://arxiv.org/abs/1307.0026}{arXiv:1307.0026 [nucl-th]}].

\bibitem{OConnell:1995fwv}
H.~B.~O'Connell, B.~C.~Pearce, A.~W.~Thomas and A.~G.~Williams,
\textit{$\rho - \omega$ mixing and the pion electromagnetic form-factor},
\href{https://www.sciencedirect.com/science/article/pii/037026939500642X}{Phys. Lett. B \textbf{354} (1995), 14-19
doi:10.1016/0370-2693(95)00642-X},
[\href{https://arxiv.org/abs/hep-ph/9503332}{arXiv:hep-ph/9503332 [hep-ph]}].


\bibitem{Cundy:2009yy}
N.~Cundy, M.~G\"{o}ckeler, R.~Horsley, T.~Kaltenbrunner, A.~D.~Kennedy, Y.~Nakamura, H.~Perlt, D.~Pleiter, P.~E.~L.~Rakow and A.~Sch\"{a}fer, G.~Schierholz, A.~Schiller, H.~Stüben and J.~M.~Zanotti,
\textit{Non-perturbative improvement of stout-smeared three flavour clover fermions},
\href{https://journals.aps.org/prd/abstract/10.1103/PhysRevD.79.094507}{Phys. Rev. D \textbf{79} (2009) 094507
doi:10.1103/PhysRevD.79.094507},
[\href{https://arxiv.org/abs/0901.3302}{arXiv:0901.3302 [hep-lat]}].

\bibitem{Bietenholz:2010jr}
W.~Bietenholz, V.~Bornyakov, N.~Cundy, M.~G\"{o}ckeler, R.~Horsley, A.~D.~Kennedy, W.~G.~Lockhart, Y.~Nakamura, H.~Perlt, D.~Pleiter, G.~Schierholz, A.~Schiller, H.~Stüben and J.~M.~Zanotti,
\textit{Tuning the strange quark mass in lattice simulations}
\href{https://www.sciencedirect.com/science/article/pii/S0370269310006702}{Phys. Lett. B \textbf{690} (2010) 436-441
doi:10.1016/j.physletb.2010.05.067},
[\href{https://arxiv.org/abs/1003.1114}{arXiv:1003.1114 [hep-lat]}].

\bibitem{Bietenholz:2011qq}
W.~Bietenholz, V.~Bornyakov, M.~G\"{o}ckeler, R.~Horsley, W.~G.~Lockhart, Y.~Nakamura, H.~Perlt, D.~Pleiter, P.~E.~L.~Rakow and G.~Schierholz, A.~Schiller, T.~Streuer, H.~Stüben, F.~Winter and J.~M.~Zanotti,
\textit{Flavour blindness and patterns of flavour symmetry breaking in lattice simulations of up, down and strange quarks}
\href{https://journals.aps.org/prd/abstract/10.1103/PhysRevD.84.054509}{Phys. Rev. D \textbf{84} (2011) 054509
doi:10.1103/PhysRevD.84.054509},
[\href{https://arxiv.org/abs/1102.5300}{arXiv:1102.5300 [hep-lat]}].

\bibitem{Chambers:2017hmh}
A.~Chambers, J.~Dragos, R.~Horsley, Y.~Nakamura, H.~Perlt, D.~Pleiter, P.~Rakow, G.~Schierholz, A.~Schiller and K.~Somfleth, H.~St\"{u}ben, R.~Young and J.~Zanotti,
\textit{Hadron Structure from the Feynman-Hellmann Theorem},
\href{https://pos.sissa.it/256/168/}{PoS \textbf{LATTICE2016} (2017) 168
doi:10.22323/1.256.0168}

\bibitem{CSSM:2014uyt}
A.~J.~Chambers, R.~Horsley, Y.~Nakamura, H.~Perlt, D.~Pleiter, P.~E.~L.~Rakow, G.~Schierholz, A.~Schiller, H.~St\"{u}ben, R.~D.~Young and J.~M.~Zanotti,
\textit{Feynman-Hellmann approach to the spin structure of hadrons},
\href{https://journals.aps.org/prd/abstract/10.1103/PhysRevD.90.014510}{Phys. Rev. D \textbf{90} (2014) 1, 014510
doi:10.1103/PhysRevD.90.014510},
[\href{https://arxiv.org/abs/1405.3019}{arXiv:1405.3019 [hep-lat]}].



\bibitem{QCDSF:2012mkm}
R.~Horsley, R.~Millo, Y.~Nakamura, H.~Perlt, D.~Pleiter, P.~E.~L.~Rakow, G.~Schierholz, A.~Schiller, F.~Winter and J.~M.~Zanotti,
\textit{A Lattice Study of the Glue in the Nucleon},
\href{https://www.sciencedirect.com/science/article/pii/S0370269312007447}{Phys. Lett. B \textbf{714} (2012), 312
doi:10.1016/j.physletb.2012.07.004},
[\href{https://arxiv.org/abs/1205.6410}{arXiv:1205.6410 [hep-lat]}].


\bibitem{Haar:2017ubh}
T.~R.~Haar, Y.~Nakamura and H.~St\"uben,
\textit{An update on the BQCD Hybrid Monte Carlo program},
\href{https://www.epj-conferences.org/articles/epjconf/abs/2018/10/epjconf_lattice2018_14011/epjconf_lattice2018_14011.html}{EPJ Web Conf. \textbf{175} (2018), 14011
doi:10.1051/epjconf/201817514011},
[\href{https://arxiv.org/abs/1711.03836}{arXiv:1711.03836 [hep-lat]}].

\bibitem{Edwards:2004sx}
R.~G.~Edwards and B.~Joo,
\textit{The Chroma software system for lattice QCD},
\href{https://www.sciencedirect.com/science/article/pii/S0920563204007455}{Nucl. Phys. B Proc. Suppl. \textbf{140} (2005), 832
doi:10.1016/j.nuclphysbps.2004.11.254},
[\href{https://arxiv.org/abs/hep-lat/0409003}{arXiv:hep-lat/0409003 [hep-lat]}].





\end{thebibliography}
\end{document}